\documentclass[conference,draftcls,onecolumn]{IEEEtran}	% SITE

\usepackage[latin1]{inputenc}
\usepackage[T1]{fontenc}
\usepackage[english]{babel}
\usepackage{amssymb,amsmath,amsfonts}
\usepackage{times}
\usepackage{graphicx}
\usepackage{color}
\usepackage{verbatim}
\usepackage{cite}

\usepackage{subfig}	
\usepackage{pgf}	
\usepackage{tikz}	
\usepackage{pgfplots}
\usetikzlibrary{shapes.geometric}
\usetikzlibrary{decorations}
\usetikzlibrary{patterns}
\usetikzlibrary{positioning}
\usetikzlibrary{backgrounds}

\usepackage{stmaryrd}

\usepackage{cancel}
\usepackage[normalem]{ulem}

\newcommand{\red}[1]{{\color{red} #1}}
\newcommand{\green}[1]{{\color{green} #1}}

\colorlet{vert}{green!70!black}

\newtheorem{definition}	{Definition}
\newtheorem{theorem}	{Theorem}

\newtheorem{proposition}{Proposition}

\newtheorem{remark}		{Remark}

\newcommand{\cA}{{\mathcal A}}
\newcommand{\cB}{{\mathcal B}}
\newcommand{\cE}{{\mathcal E}}
\newcommand{\cN}{{\mathcal N}}
\newcommand{\cQ}{{\mathcal Q}}
\newcommand{\cR}{{\mathcal R}}

\newcommand{\cT}{{\mathcal T}}
\newcommand{\cU}{{\mathcal U}}
\newcommand{\cV}{{\mathcal V}}
\newcommand{\cX}{{\mathcal X}}
\newcommand{\cY}{{\mathcal Y}}
\newcommand{\cZ}{{\mathcal Z}}

\newcommand{\bN}{{\mathbb N}}
\newcommand{\bR}{{\mathbb R}}

\newcommand{\bE}{{\mathbb E}}

\newcommand{\Var}[1]{\text{Var}\left[#1\right]}

\newcommand{\mkv}{-\!\!\!\!\minuso\!\!\!\!-}

\newcommand{\typ}[1]{T_\varepsilon^n(#1)}

\providecommand{\norm}[1]{\lVert#1\rVert}

\newcommand{\lessnoisy}[1]{\succeq_{\scriptscriptstyle #1}}

\newcommand{\indpt}{\perp\!\!\!\perp}

\usepackage{hyperref}			% SITE
\IEEEoverridecommandlockouts	% pour faire \thanks

\begin{document}

\title{Hybrid Digital/Analog Schemes for Secure Transmission with Side Information} 

\author{
\IEEEauthorblockN{
	Joffrey Villard\IEEEauthorrefmark{1},
	Pablo Piantanida\IEEEauthorrefmark{1} and
	Shlomo Shamai (Shitz)\IEEEauthorrefmark{2}
}
\\
\IEEEauthorblockA{
\begin{tabular}{cc}
	\IEEEauthorrefmark{1}	Department of Telecommunications	&
	\IEEEauthorrefmark{2}	Department of Electrical Engineering	\\
	SUPELEC										&	Technion - Israel Institute of Technology\\	
	91192 Gif-sur-Yvette, France				&	Technion city, Haifa 32000, Israel\\
	Email: \{joffrey.villard, pablo.piantanida\}@supelec.fr &
	Email: sshlomo@ee.technion.ac.il	
\end{tabular}
}

\thanks{The work of J. Villard is supported by DGA (French Armament Procurement Agency). This research is partially supported by the FP7 Network of Excellence in Wireless COMmunications NEWCOM++.}
}

\date{September 2011} % SITE 

\maketitle
 
\renewcommand{\leftmark}{\MakeUppercase{to be presented at ITW 2011}}		% SITE
\renewcommand{\rightmark}{} 												% SITE
 
%--------------------------------------------------------------------
\begin{abstract}
Recent results on source-channel coding for secure transmission show that separation holds in several cases under some less-noisy conditions.
However, it has also been proved through a simple counterexample that pure analog schemes can be optimal and hence outperform digital ones.
According to these observations and assuming matched-bandwidth, we present a novel hybrid digital/analog scheme that aims to gather the advantages of both digital and analog ones. In the quadratic Gaussian setup when side information is only present at the eavesdropper, this strategy is proved to be optimal. Furthermore, it outperforms both digital and analog schemes and cannot be achieved via time-sharing. An application example to binary symmetric sources with side information is also investigated.
\end{abstract}

%==============================================================================
\section{Introduction}

\begin{figure}[t]
\centering
\begin{tikzpicture}[xscale=.95,yscale=.9]
	\node						(A) 	at (0,0) 	{\small $A^n$};
	\node[rectangle,draw,blue]	(alice)	at (1.5,0) 	{\small Alice};
	\node[rectangle,draw,vert] 	(canal)	at (4,0)	{\small \begin{tabular}{c}DMC\\ $\!\!\!\!\! p(yz|x)\!\!\!\!\!$\end{tabular}};
	
	\node						(B) 	at (0,1.5) 	{\small $B^n$};
	\node[rectangle,draw,blue]	(bob) 	at (6.5,.5)	{\small Bob};
	\node[right,blue]			(hatA) 	at (7.3,.5)	{\small $\hat A^n: d(A^n,\hat A^n)\lesssim D$};
	
	\node[rectangle,draw,red]	(eve) 	at (6.5,-.5){\small Eve};
	\node	(E) 		at (0,-1.5)					{\small $E^n$};
	\node[right,red]			(D) 	at (7.3,-.5){\small $\frac1n H(A^n|E^n Z^n)\gtrsim\Delta$};
		
	\draw[->]	(A)			to	(alice);
	\draw[->]	(B)			-|  (bob);
	\draw[->]	(E)			-|	(eve);
	
	\draw[->,vert]	(alice)		to node[above]{\small $X^n$} (canal);
	
	\draw[->,vert]	(canal.east|-bob)	to node[above]{\small $Y^n$} (bob);
	\draw[->,vert]	(canal.east|-eve)	to node[below]{\small $Z^n$} (eve);
	
	\draw[->,blue]	(bob)	to (hatA);
	\draw[->,red]	(eve)	to (D);
\end{tikzpicture}
\caption{Secure transmission with side information.}
\label{fig:schema}
\end{figure}
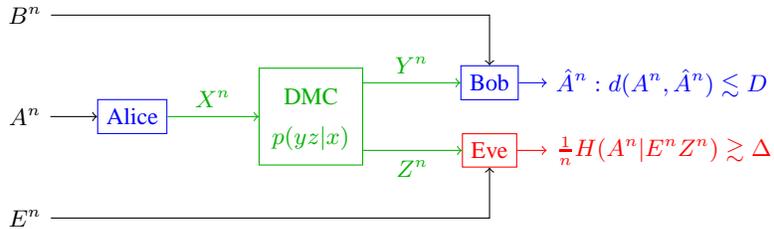

The setup of source-channel coding for secure transmission consists of three nodes that measure an analog source as a function of time.
One of them (referred to as Alice) wishes to transmit a compressed version of its observation to a second node (referred to as Bob) through a noisy (or wireless) channel.
In addition, Bob can use his own observation as side information to decode the received message and refine his estimate of Alice's source.
The third node (referred to as Eve) is an eavesdropper \emph{i.e.}, a node that can listen to the messages sent by Alice through another noisy channel, as shown in Fig.~\ref{fig:schema}. Considering Eve as an untrusted node, Alice wishes to leak the smallest amount of information about her source.

Among some major information-theoretic issues, the above scenario involves the notion of secrecy (and its application to source and channel coding), source coding with side information, as well as joint source/channel coding for transmission of sources over noisy channels.
The information-theoretic notion of secrecy was introduced by Shannon~\cite{shannon1949communication} and used for secure communication over noisy channels by Wyner~\cite{wyner1975wire}, who introduced the wiretap channel, which was further extended by Csisz\'ar and K\"orner~\cite{csiszar1978broadcast}.
On the other hand, source coding with side information has been introduced by Slepian and Wolf~\cite{slepian1973noiseless}, and Wyner and Ziv~\cite{wyner1976rate}.
Recent results~\cite{prabhakaran2007secure,villard2010secure} consider such settings with an additional eavesdropper that must be kept as ignorant as possible of the transmitted source.
Most of the existent work separately consider channel or source coding for secure transmission or compression. However, unlike simple point-to-point communication problems, there is no general result of separation for multiterminal settings under security constraints. Recent work~\cite{merhav2008shannon,villard2011secureb} about source-channel coding for secure transmission shows that separation holds in several cases under some less-noisy conditions. Whereas a simple counterexample in~\cite{villard2011secureb} showed that a pure analog scheme can improve digital schemes while being optimal. This observation motivates us to investigate hybrid digital/analog schemes for such setting, which by taking advantage of both analog and digital strategies may yield better performance. Indeed, such schemes have already been proved useful for point-to-point problems \emph{e.g.}, to handle SNR mismatch (while they can perform as good as digital or analog ones at the true SNR)~\cite{mittal2002hybrid,wilson2010joint}, as well as for some multiterminal settings~\cite{gunduz2008wyner-ziv,lim2010lossy}.

%::::::::::::::::::::::::::::::::::::::
%::::::::::::::::::::::::::::::::::::::

In this paper, we consider the setup of joint source-channel coding for secure transmission of a source over a noisy channel with an eavesdropper, and in the presence of side information at the receiving terminals. We restrict our attention to the matched-bandwidth case \emph{i.e.}, one channel use is allowed per source symbol, as depicted in Fig.~\ref{fig:schema}. Our main goal is to understand how Alice can take simultaneous advantage of both statistical differences among side informations and channels. Section~\ref{sec:def_prior} recalls some prior results that yield an inner bound (digital scheme), and a general outer bound on the distortion-equivocation region.
A novel hybrid digital/analog scheme and its single-letter inner bound is derived in Section~\ref{sec:hybrid}. Transmission of a Gaussian source over a Gaussian wiretap channel with side information  (resp. binary source over a type-II wiretap channel) is studied in Section~\ref{sec:gaussian} (resp. Section~\ref{sec:binary}).

%------------------------------------------------------------------------------
\subsection*{Notations}
For any sequence~$(x_i)_{i\in\bN^*}$, notation $x^n$
stands for the collection $(x_1,x_2,\dots, x_n)$.
Entropy is denoted by $H(\cdot)$, and mutual information by $I(\cdot;\cdot)$.
Let $X$, $Y$ and $Z$ be three random variables on some alphabets with probability distribution~$p$.
If $p(x|y,z)=p(x|y)$ for each $x,y,z$, then they form a Markov chain, which is denoted by $X\mkv Y\mkv Z$.
The set of nonnegative real numbers is denoted by $\bR_+$.
For each $x\in\bR$, notation $[x]_+$ stands for $\max(0;x)$.
Notation $h_2$ stands for the binary entropy function.
The binary exclusive-or operator is denoted by $\oplus$.

% The Gaussian distribution with mean $\mu$ and variance $\sigma^2$ is denoted by $\cN(\mu,\sigma^2)$.
% The Bernoulli distribution of parameter $u$ is denoted by $\cB(u)$.

%==============================================================================
\section{Problem Definition and Prior Results}
\label{sec:def_prior}

%------------------------------------------------------------------------------
\subsection{Problem Definition}

Let $\cA$, $\cB$, $\cE$, $\cX$, $\cY$, and $\cZ$ be six finite sets. 
Alice, Bob, and Eve observe the sequences of random variables 
$(A_i)_{i\in\bN^*}$, $(B_i)_{i\in\bN^*}$, and $(E_i)_{i\in\bN^*}$,
respectively, which take values on $\cA$, $\cB$, and $\cE$, resp.
For each $i\in\bN^*$, the random variables $A_i$, $B_i$, and $E_i$
are distributed according to the joint distribution $p(abe)$ on
$\cA\times\cB\times\cE$.
Moreover, they are independent across time $i$.
Alice can also communicate with Bob and Eve through a discrete memoryless channel $p(yz|x)$ with input $X$ on $\cX$, and outputs $Y$, $Z$ on $\cY$, $\cZ$, respectively. Let $d : \cA\times\cA \to [0\,;d_{\max}]$ be a finite distortion measure
\emph{i.e.}, $0\leq d_{\max} < \infty$.
Denote by $d$ the component-wise mean distortion on $\cA^n\times\cA^n$
\emph{i.e.}, for each $a^n,b^n\in\cA^n$, $d(a^n,b^n) = \frac1n\,\sum_{i=1}^n d(a_i,b_i)$.

\begin{definition}[Code and achievability]
\label{def:achievability}
An $n$-code for source-channel coding in this setup is defined by
\begin{itemize}
\item A (stochastic) encoding function at Alice $F : \cA^n \to \cX^n$, defined by some transition probability $P_{X^n|A^n}(\cdot|\cdot)$,
\item A decoding function at Bob $g : \cB^n\times\cY^n \to \cA^n$.
\end{itemize}
A tuple $(D,\Delta)\in\bR_+^2$ is \emph{achievable} if,
for any $\varepsilon>0$, there exists an $n$-code $(F,g)$ s.t.:
\begin{IEEEeqnarray*}{rCl}
\bE\big[ d(A^n,g(B^n,Y^n)) \big]	&\leq& D+\varepsilon \ ,\\
\dfrac1n\,H(A^n|E^n Z^n) 			&\geq& \Delta-\varepsilon \ ,
\end{IEEEeqnarray*}
with channel input $X^n$ as the output of the encoder $F(A^n)$.
The set of all achievable tuples is denoted by $\cR^*$
and is referred to as the \emph{distortion-equivocation region}.
\end{definition}

%------------------------------------------------------------------------------
\subsection{Prior Results}
\label{sec:prior}

Recent results~\cite{villard2011secureb} provide inner and outer bounds on region $\cR^*$ which do not match in general.
In the matched-bandwidth case, the inner bound $\cR_\text{dig}$ is defined by the next theorem.

\begin{theorem}[Digital Scheme~\cite{villard2011secureb}]
\label{th:digital}
The set of all tuples $(D,\Delta)\in\bR_+^2$ such that there exist RVs $U$, $V$, $Q$, $T$, $X$ on finite sets $\cU$, $\cV$, $\cQ$, $\cT$, $\cX$ respectively, with joint distribution $p(uvqtabexyz) = p(u|v)p(v|a)p(abe)\,p(q|t)p(tx)\linebreak p(yz|x)$, and a function $\hat A : \cV\times\cB \to \cA$, verifying the following inequalities, is achievable:
\begin{IEEEeqnarray*}{l}
I(U;A|B) \leq I(Q;Y) 										\ ,\\
I(V;A|B) \leq I(T;Y) 										\ ,\\
D 		 \geq \bE\big[d(A,\hat A(V,B))\big] 				\ ,\\
\Delta	 \leq  H(A|UE) - \Big[ I(V;A|UB) 
				- \Big( I(T;Y|Q) - I(T;Z|Q) \Big) \Big]_+	\ .
\end{IEEEeqnarray*}
\end{theorem}

The achievability of the above region follows by combining secure source coding~\cite{villard2010secure} with coding for broadcast channels with common message~\cite{csiszar1978broadcast}. This results in two independent (but not stand-alone) source and channel components, yielding statistically independent source and channel variables \emph{i.e.}, ``operational'' separation holds. Such a scheme will be referred to as the \emph{digital scheme}. In some special cases \emph{e.g.}, when Eve has less noisy channel, or Bob has less noisy side information, this inner bound becomes tight~\cite{villard2011secureb} and traditional separation holds.
Nevertheless, when Bob has ``better'' channel and ``worse'' side information than Eve, it was observed in~\cite{villard2011secureb} through a simple example that a naive analog scheme, consisting of directly plugging the source on the channel, outperforms the digital one of Theorem~\ref{th:digital}. Furthermore, it turns to be optimal since it can be shown that it achieves the next outer bound $\cR_\text{out}$.

\begin{theorem}[Outer Bound~\cite{villard2011secureb}]
\label{th:outer}
For each achievable tuple $(D,\Delta)$, there exist random variables $U$, $V$, $Q$, $T$, $X$ on finite sets $\cU$, $\cV$, $\cQ$, $\cT$, $\cX$, respectively, and a function $\hat A : \cV\times\cB \to \cA$, such that $p(uvqtabexyz) = p(uv|a)p(abe)\,p(q|t)p(tx)p(yz|x)$, and
\begin{IEEEeqnarray*}{l}
I(V;A|B) \leq I(T;Y) 											\ ,\\
D 		 \geq \bE\big[d(A,\hat A(V,B))\big] 					\ ,\\
\Delta	 \leq  H(A|UE) - \Big[ I(V;A|B) - I(U;A|B)		
					- \Big( I(T;Y|Q) - I(T;Z|Q) \Big) \Big]_+	\ .
\end{IEEEeqnarray*} 
\end{theorem}

%==============================================================================
\section{Hybrid Coding for Secure Transmission}
\label{sec:hybrid}

Based on the observation  made in~\cite{villard2011secureb} about the usefulness of \emph{analog schemes}, we now propose a \emph{hybrid digital/analog scheme} that yields the next inner bound $\cR_\text{hyb}\subseteq \cR^*$.

%------------------------------------------------------------------------------
\subsection{Main Result}

\begin{theorem}[Hybrid Scheme]
\label{th:hybrid}
The set of all tuples $(D,\Delta)$ in $\bR_+^2$ such that there exist RVs $U$, $V$, $X$ on finite sets $\cU$, $\cV$, $\cX$, with joint PD $p(uvabexyz) = p(u|v)p(vx|a)\,p(abe)\,p(yz|x)$, $x=x(v,a)$, and a function $\hat A : \cV\times\cB\times\cY \to \cA$, verifying the following inequalities, is achievable:
\begin{IEEEeqnarray}{l}
I(U;A)		\leq I(U;BY) 										\ ,	\label{eq:hybrid1}\\
I(V;A|U)	\leq I(V;BY|U)										\ ,	\label{eq:hybrid2}\\[.1cm]
D			\geq \bE\big[d(A,\hat A(V,B,Y))\big]				\ ,	\label{eq:hybrid3}\\[.1cm]
\Delta		\leq  H(A|UE) - I(V;A|U) - I(X;Z|UE) 				
				+ \min\Big\{ I(V;BY|U) \ ;\ I(V;AZ|U) \Big\}	\ .	\label{eq:hybrid4}
\end{IEEEeqnarray}
\end{theorem}

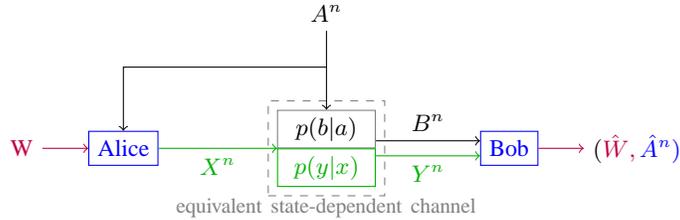
\begin{figure}
\centering
\begin{tikzpicture}[scale=.9]
	\tikzstyle{trans}=[text centered,inner sep=0,minimum height=.5cm,minimum width=1.3cm,rectangle,draw,solid];
	
	\node[purple]				(noise)		at (  0,  0)		{\small W};
	\node[rectangle,draw,blue]	(alice)		at (1.5,  0)	 	{\small Alice};
	
	\node[matrix,draw=gray,dashed] at (4.5,  0){
		\node[trans]		(probB)		{\small $p(b|a)$}; \\
		\node[trans,vert]	(canal)		{\small $p(y|x)$}; \\
	};	
	
	\node[below=8pt,gray]	at (canal)								{\footnotesize equivalent state-dependent channel};
	
	\node						(A) 		at (4.5,  2) 		{\small $A^n$};
	\node[rectangle,draw,blue]	(bob) 		at (7.2,  0)	 	{\small Bob};
	\node[right]				(hatA) 		at (8.3,  0)		{\small $(\color{purple}{\hat W},\color{blue}{\hat A^n})$};
		
	\draw[->,purple]	(noise) to (alice);
	
	\draw[->]			(A)			to (probB);
	\draw[->]			(A |- 0,1.2)-| (alice);
	
	\node (bg) at (probB.south east) [above=.1] {};
	\node (bd) at (bob.west) 		 [above=.1] {};
	\draw[->]			(bg.south)	to node[above]{\small $B^n$} (bd.south);
	
	\draw[->,vert]		(alice)		to node[below]{\small $X^n$} (canal.north west);
	
	\node (yg) at (canal.north east) [below=.1] {};
	\node (yd) at (bob.west) 		 [below=.1] {};
	\draw[->,vert]		(yg.north)	to node[below]{\small $Y^n$} (yd.north);
	
	\draw[->,purple]	(bob)	to (hatA);
\end{tikzpicture}
\caption{Alice and Bob as a communication system with state-dependent channel and CSIT.}
\label{fig:gelfand}
\end{figure}

Channels $A\mapsto B$ and $X\mapsto Y$ can be viewed together as a state-dependent channel with input $X$, state $A$ and output $(B,Y)$.
In this perspective, Alice and Bob form a communication system with channel state information known at the transmitter (CSIT), as depicted in Fig.~\ref{fig:gelfand}.
Roughly speaking, the proposed strategy consists in sending independent digital random noise $W$ using a Gelfand-Pinsker code~\cite{gel'fand1980coding} for this equivalent state-dependent channel. Two auxiliary RVs $(U,V)$ constitute the descriptions of $A$.
Digital random noise $W$ here helps to secure description $V$ against Eve.
As in the classical wiretap channel~\cite{csiszar1978broadcast,liang2009information}, its rate satisfies some constraint that allows to characterize the equivocation rate at Eve. Finally, the auxiliary RV $U$ is a subpart of $V$ that can be seen as a \emph{common} message which is considered to be known at Eve, as shown by the term $H(A|UE)$ in~\eqref{eq:hybrid4}.

%::::::::::::::::::::::::::::::::::::::
%::::::::::::::::::::::::::::::::::::::

%------------------------------------------------------------------------------
\subsection{Sketch of Proof}

We provide a short sketch of proof of Theorem~\ref{th:hybrid}. 
Details will be given in an extended version of this paper.
Let $R_1, R_2, R_f\in\bR_+^*$ and assume that a local (independent and uniformly distributed) source with rate $R_f$ is available at Alice.

% - - - - - - - - - - - - - - - - - - - - - - - - - - - - - - - - - - - - - - -
\subsubsection{Codebook generation}

Randomly pick $2^{nR_1}$ sequences $u^n(r_1)$ from $\typ{U}$.
Then, for each  $u^n(r_1)$, randomly pick $2^{n(R_2+R_f)}$ sequences $v^n(r_1,r_2,r_f)$ from $\typ{V|u^n(r_1)}$.

% - - - - - - - - - - - - - - - - - - - - - - - - - - - - - - - - - - - - - - -
\subsubsection{Encoding}

Assume that source sequence $A^n$ and random noise $r_f$ are produced at Alice.
Look for the first codeword $u^n(r_1)$ such that $(u^n(r_1),A^n)\in\typ{U,A}$.
Then look for the first codeword $v^n(r_1,r_2,r_f)$ such that $(v^n(r_1,r_2,r_f),A^n)\in\typ{V,A|u^n(r_1)}$ and send $X^n$, defined by the component-wise relation 
\[
X_i \triangleq x(v_i(r_1,r_2,r_f),A_i) \ .
\] 
By standard arguments, it can be proved that these two steps succeed with high probability if $R_1>I(U;A)$ and $R_2>I(V;A|U)$.

% - - - - - - - - - - - - - - - - - - - - - - - - - - - - - - - - - - - - - - -
\subsubsection{Decoding}

Assume that Bob observes $B^n$ and receives $Y^n$ from Alice.
Look for the unique codeword $u^n(r_1)$ such that $(u^n(r_1),B^n,Y^n) \in \typ{U,B,Y}$.
Then look for the unique codeword $v^n(r_1,r_2,r_f)$ such that $(v^n(r_1,r_2,r_f),B^n,Y^n) \in \typ{V,B,Y|u^n(r_1)}$.
Compute the estimate $g(B^n,Y^n)\in\cA^n$ using the component-wise relation
\[
g_i(B^n,Y^n) \triangleq \hat A(v_i(r_1,r_2,r_f),B_i,Y_i)\ .
\]

As noted in~\cite{lim2010lossy}, the conventional random coding proof technique does not apply here, and the decoding error probability must be carefully handled.  In the proposed joint coding scheme, a single codebook plays both roles of source and channel codebooks. For a given source sequence $a^n$, the indices $(r_1,r_2)$ thus depend on the entire codebooks, and the averaging over the set of all possible codebooks cannot be performed in the usual way.
Similarly to~\cite{lim2010lossy}, it is not difficult to show that these two decoding steps succeed with high probability if $R_1<I(U;BY)$ and $R_2+R_f<I(V;BY|U)$.

% - - - - - - - - - - - - - - - - - - - - - - - - - - - - - - - - - - - - - - -
\subsubsection{Distortion at Bob}

Provided the above constraints are verified, Bob can decode codeword $v^n(r_1,r_2,r_f)$ with an arbitrarily small probability of error, and compute estimate $g(Y^n,B^n)$ of $A^n$.
Using standard properties of typical sequences, it can be easily proved that the mean distortion of this estimate approaches $\bE[ d(A,\hat A(V,B,Y)) ]$.

% - - - - - - - - - - - - - - - - - - - - - - - - - - - - - - - - - - - - - - -
\subsubsection{Equivocation rate at Eve}

The equivocation at Eve can be divided in ``source'' and ``channel'' terms. 
Each one is studied using standard properties of typical sequences, and the arguments of~\cite{villard2010secure} and~\cite[Section~2.3.\emph{Step~3}]{liang2009information}. The next constraint which ensures that Eve decodes $v^n(r_1,r_2,r_f)$ from $(r_1,A^n,Z^n)$ is required to characterize the equivocation
\[
R_2 + R_f < I(V;AZ|U) \ .
\]
After some derivations, we can eventually prove that
\begin{IEEEeqnarray*}{rCl}
\frac1n H(A^n | E^n Z^n)
	\geq 	H(A|UE) - I(X;Z|UE) + R_f - \varepsilon	\ .
\end{IEEEeqnarray*}

% - - - - - - - - - - - - - - - - - - - - - - - - - - - - - - - - - - - - - - -
\subsubsection{End of Proof}

Gathering the above inequalities and performing Fourier-Motzkin elimination prove Theorem~\ref{th:hybrid}.

%------------------------------------------------------------------------------
\subsection{Special Cases}

% - - - - - - - - - - - - - - - - - - - - - - - - - - - - - - - - - - - - - - -
\subsubsection{Analog schemes}
\label{sec:specAnalog}
The proposed scheme can reduce to a pure analog one (as the simple one of~\cite[Section~VI]{villard2011secureb}).
Hence  $\cR_\text{hyb}$ contains tuples that may not be in $\cR_\text{dig}$: $\cR_\text{hyb}\not\subset\cR_\text{dig}$.

% - - - - - - - - - - - - - - - - - - - - - - - - - - - - - - - - - - - - - - -
\subsubsection{Digital schemes}
\label{sec:specDigital}

By defining the variables in Theorem~\ref{th:hybrid} as pairs of independent source and channel components, we can obtain the structure of those in Theorem~\ref{th:digital}, but such variables do not verify all inequalities and thus $\cR_\text{dig}\not\subset\cR_\text{hyb}$.

% - - - - - - - - - - - - - - - - - - - - - - - - - - - - - - - - - - - - - - -
\subsubsection{Wiretap channel}
Choosing independent source and channel variables with appropriate rates, region $\cR_\text{hyb}$ reduces to the achievable region for the wiretap channel~\cite[Eq.~(2.6)]{liang2009information}.

%==============================================================================
\section{Secure Transmission of a Gaussian Source over a Gaussian Wiretap Channel}
\label{sec:gaussian}

%------------------------------------------------------------------------------
\subsection{System Model}

In this section, we consider the transmission of a Gaussian source over a Gaussian wiretap channel.
More precisely, the source at Alice $A$ is standard Gaussian, and observations at Bob and Eve are the outputs of independent additive white Gaussian noise (AWGN) channels with input $A$ and respective noise powers $P_B$ and $P_E$.
Communication channels from Alice to Bob and Charlie are AWGN channels with respective noise powers $P_Y$ and $P_Z$. 
The average input power of this channel is limited to $P$. 

Euclidean distance on $\bR$ is used to measure distortion at Bob ($d(a,b) = (a-b)^2$, for each $a,b\in\bR$). 
Differential entropy $h(\cdot)$ measures uncertainty yielding equivocation rates $\Delta\in\bR$. 
We also introduce quantity $D_E\triangleq 2^{2\Delta}/(2\pi e)$, which provides a lower bound on the minimum mean-square error of any estimator of $A$ at Eve.

\begin{definition}[Achievability]
In this section, a tuple $(D,D_E)\in{\bR_+^*}^2$ is said to be \emph{achievable} if,
for any $\varepsilon>0$, there exists an $n$-code $(F,g)$ s.t.:
\begin{IEEEeqnarray*}{rCl}
\bE\big[ \norm{A^n - g(B^n,Y^n)}^2 \big]	&\leq& D+\varepsilon 		\ ,\\[1mm]
\dfrac1n\,h(A^n|E^n Z^n) 					&\geq& \frac12\log\left( 2\pi e\,D_E \right)-\varepsilon	\ ,\\[1mm]
\frac1n \sum_{i=1}^n \bE\big[ X_i^2 \big]	&\leq& P+\varepsilon 		\ ,
\end{IEEEeqnarray*}
with channel input $X^n$ as the output of the encoder $F(A^n)$.
\end{definition}
\pagebreak

Although Theorems~\ref{th:digital}--\ref{th:hybrid} are stated and proved for finite alphabet, we take the liberty to use their statements as inner regions also for this case. The involved joint PDs should now also verify condition $\Var{X} \leq P$. The corresponding regions will be denoted with an additional $\cdot^P$ \emph{i.e.}, $\cR_\text{dig}^P$, $\cR_\text{out}^P$ and $\cR_\text{hyb}^P$. 

Notice that due to the Gaussian additive noises, and depending on the relative values of $P_B$, $P_E$ (resp. $P_Y$, $P_Z$), one side information (resp. one channel) is a stochastically degraded version of the other. There exist four different cases and, from known results (see Section~\ref{sec:prior}), separation holds for three of them, as summarized in Table~\ref{tab:combin}. For instance, the case when Bob has ``better'' channel ($P_Y < P_Z$) and ``worse'' side information ($P_B > P_E$) than Eve is still open. We next propose a hybrid digital/analog scheme based on Theorem~\ref{th:hybrid} that turns to be optimal when $P_Y < P_Z$ and $P_B\to\infty$.

%::::::::::::::::::::::::::::::::::::::
\begin{table}
\renewcommand{\arraystretch}{1.4}
\centering
\begin{tabular}{c||c|c}
				&	$P_B \leq P_E$								
				&	$P_B >    P_E$			\\\hline\hline
$P_Y <    P_Z$	&	\green{\checkmark}
				& 	\red{\bf ?}				\\\hline
$P_Y \geq P_Z$	&	\green{\checkmark}
				&	\green{\checkmark}
\end{tabular}
\caption{Cases where $\cR_\text{dig}$ is tight and separation holds.}
\label{tab:combin}
\end{table}
%::::::::::::::::::::::::::::::::::::::

%------------------------------------------------------------------------------
\subsection{Hybrid Coding}
\label{sec:gaussHybrid}

In this section, we consider hybrid coding with RVs $U$, $V$ and $X$ of Theorem~\ref{th:hybrid} defined as $U = \emptyset$ and 
\begin{IEEEeqnarray}{rCl}
V &=& \alpha A + \gamma N							\ ,\label{eq:defV}\\
X &=& \sqrt{P} \big( \beta A - \gamma N \big)		\ ,\label{eq:defX}
\end{IEEEeqnarray}
where $\gamma = \sqrt{1 - \beta^2}$,
and $N\sim\cN(0,1)$ is a standard Gaussian random variable independent of $A$.
Note that $X\sim\cN(0,P)$ writes as a deterministic function of $A$ and $V$:
\[
X = \sqrt{P}\big( (\alpha+\beta) A - V \big) \ .
\]
$\hat A$ is defined as the MMSE estimator of $A$ from $(V,Y)$.

\begin{figure}
\centering
\begin{tikzpicture}
	\tikzstyle{add}=[inner sep=0pt,minimum width=.5cm,draw,circle];
	\tikzstyle{mult}=[fill=lightgray,draw,isosceles triangle,inner sep=1pt];
	\tikzstyle{codeword}=[draw,minimum width=.8cm];
	
	% - - - - - - - - - - - - - - - - -	
	\node[matrix,draw,every node/.style={codeword}] (codebook) at (0,0){
		\node{};&		\node{};&		\node{};& 			\node{}; 	\\
		\node{};& 		\node{};& 		\node{};& 			\node{}; 	\\
		\node{};& 		\node{};& 		\node{};& 			\node{}; 	\\
		\node{};& 		\node{};& 		\node[fill=gray]{};&\node{}; 	\\
		\node{};&		\node{};& 		\node{};& 			\node{}; 	\\
	};
	\node[left=1]	(A)		at (codebook.west) 	{$A^n$};
	\node[above=.5]	(r) at (codebook.north) 	{$r_f\sim\cU_{\{1,\dots,2^{nR_f}\}}$};
	
	\node[mult,below=1.4]		(mult)	at (codebook)	{$\times$};
	\node[below]	at (mult.south)	{$(\alpha+\beta)$};
	
	\node[add,right=1]			(add)	at (codebook.east)	{\large $+$};
	\node[left=.05]	at (add.south west) {\footnotesize $-$};
	\node[below=.05]at (add.south west) {\footnotesize $+$};
	
	\node[mult,right=.7]	(multP)	at (add.east)	{$\times$};
	\node[below]	at (multP.south){$\sqrt{P}$};
	
	\node[right=1.2]	(X) at (multP) {$X^n$};
	
	\draw[->]		(A)	to (codebook);
	\draw[->]		(codebook.west) ++ (-.4,0) |- (mult);
	
	\draw[->]		(r)	to (codebook);
	\draw[->]		(codebook)	to node[above]{$v^n$} (add);
	\draw[->]		(mult)	-| (add);
	
	\draw[->]		(add)	to (multP);
	\draw[->]		(multP)	to (X);
\end{tikzpicture}
\caption{Hybrid digital/analog scheme for secure transmission of a Gaussian source over a Gaussian wiretap channel.}
\label{fig:hybrid}
\vspace{-2mm}
\end{figure}
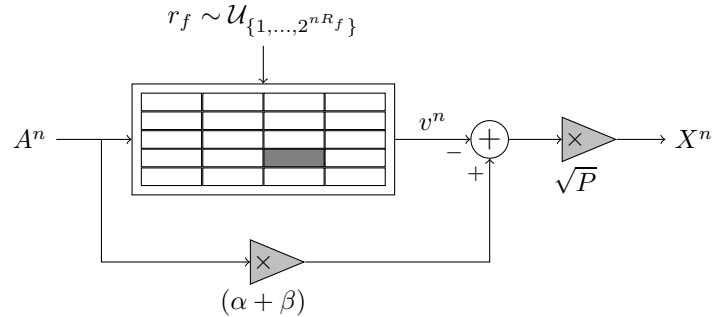

The hybrid digital/analog scheme of Section~\ref{sec:hybrid} with the above variables reduces to the one depicted in Fig.~\ref{fig:hybrid}.

%------------------------------------------------------------------------------
\subsection{Special Case: $P_Y < P_Z$, $P_B\to\infty$}
\label{sec:gaussNoB}

From now on, we focus on the unsolved case (represented by ``\red{\bf ?}'' in Table~\ref{tab:combin}), where 
 $P_Y < P_Z$. Then, if Bob does not have any side information $B=\emptyset$ (or equivalently $P_B\to\infty$): 

\begin{itemize}
\item The hybrid digital/analog scheme of Section~\ref{sec:gaussHybrid} is optimal and yields Theorem~\ref{th:gauss0Hybrid}.
\item The digital scheme of~\cite{villard2011secureb} is strictly sub-optimal, as shown by Proposition~\ref{prop:gauss0Isit} and Fig.~\ref{fig:DE=fD}.
\end{itemize}

\begin{theorem}[Gaussian sources]
\label{th:gauss0Hybrid}
If $P_Y < P_Z$ and $B=\emptyset$, a tuple $(D,D_E)\in{\bR_+^*}^2$ is achievable if and only if
\begin{IEEEeqnarray*}{rCl}
D 	&\geq& \frac1{1+\frac{P}{P_Y}}	\ ,\\
D_E	&\leq& \frac1{\max\left\{ 1 \,;\, \frac1D \cdot \frac{1+\frac{P}{P_Z}}{1+\frac{P}{P_Y}} \right\} +\frac1{P_E}}\ .
\end{IEEEeqnarray*}
\end{theorem}

%::::::::::::::::::::::::::::::::::::::
%::::::::::::::::::::::::::::::::::::::

\begin{IEEEproof}
The converse proof relies on the (conditional) entropy power inequality and is omitted here. The direct part follows after Theorem~\ref{th:hybrid} by choosing in the hybrid scheme of Section~\ref{sec:gaussHybrid}, for any distortion level $D\in\left[	\frac1{1+\frac{P}{P_Y}} , \frac{1+\frac{P}{P_Z}}{1+\frac{P}{P_Y}} \right]$:
\begin{IEEEeqnarray*}{rCl}
\alpha	&\triangleq&  \frac	
			{ \beta	+ \gamma^2 \sqrt{\frac1{D}\left( \frac{P}{P_Y}-\frac{P}{P_Z} \right)} }
			{ 1 + \gamma^2 \frac{P}{P_Y}}	  
			- \beta									\ ,\\[1mm]
\beta	&\triangleq& \sqrt{\frac{P_Z}{P}}\sqrt{ 1 + \frac{P}{P_Z} - D \left(1+\frac{P}{P_Y}\right) } \ .
\end{IEEEeqnarray*}
\end{IEEEproof}
The next proposition provides a simple expression of region $\cR_\text{dig}^P$ \emph{i.e.}, the set of all tuples achievable by digital scheme~\cite{villard2011secureb}.

\begin{proposition}
\label{prop:gauss0Isit}
If $P_Y < P_Z$ and $B=\emptyset$, $(D,\Delta)\in\cR_\text{dig}^P$ if and only if,
for some $\mu\in\left[ \frac1{1+\frac{P}{P_Y}} ; 1 \right]$ and $D_E = \frac{2^{2\Delta}}{2\pi e}$:
\begin{IEEEeqnarray*}{rCl}
D	&\geq& \frac1{1+\frac{P}{P_Y}}													\ ,\\
D_E &\leq& \frac1{\frac1\mu + \frac1{P_E}}\cdot
			\min\left\{ 1 \,;\,
				\frac	{D\left(1+\frac{P}{P_Y}\right)}
						{1 + \mu\frac{P}{P_Z} - (1-\mu)\frac{P_Y}{P_Z}} \right\}	\ .
\end{IEEEeqnarray*}
\end{proposition}

\begin{remark}
\label{rem:digitalOpt}
If $D \geq \frac{1+\frac{P}{P_Z}}{1+\frac{P}{P_Y}}$, then $\mu=1$ is optimal in Proposition~\ref{prop:gauss0Isit}, yielding inequalities of Theorem~\ref{th:gauss0Hybrid}.
This implies that the digital scheme of~\cite{villard2011secureb} is optimal in this region.
For such distortion levels, the quantity $D_E = \frac1{1+\frac1{P_E}}\linebreak = \Var{A|E}$ is achievable, meaning that Eve cannot retrieve additional information from the communication between Alice and Bob.
\end{remark}

% - - - - - - - - - - - - - - - - - - - - - - - - - - - - - - - - - - - - - - -
\subsubsection*{Numerical Results}

\begin{figure}
\centering
\begin{tikzpicture}

% styles des différentes courbes
\tikzstyle{optimal}	=[			color=red,	thick]
\tikzstyle{analog}	=[dotted,	color=blue,	thick]
\tikzstyle{digital}	=[dashed,	color=green,thick]
\tikzstyle{limites}	=[dashed,	color=black]

% limites et labels des axes
\newcommand{\xmin}{.3}
\newcommand{\xmax}{1}
\newcommand{\ymin}{.25}
\newcommand{\ymax}{.52}

\newcommand{\xlabel}{$D$}
\newcommand{\ylabel}{$D_E$}

\pgfplotsset{every axis x label/.append style={at={(1.12,.1)}}}

% - - - - - - - - - - - - - - - - - - - - - - - - - - - - - - - - - -
% axes, avec réglages de la grille, des légendes, des labels et des échelles pour abscisses et ordonnées
\begin{axis}[xlabel={\xlabel}, ylabel={\ylabel}, 
			xmin=\xmin, xmax=\xmax, ymin=\ymin, ymax=\ymax, 
			grid=both,
			width=6.9cm]

\pgfplotsset{every axis grid/.style={dotted}} 
\pgfplotsset{every axis legend/.append style={
	cells={anchor=west},
	at={(.97,.03)},
	anchor=south east
}}

% - - - - - - - - - - - - - - - - - - - - - - - - - - - - - - - - - -
% tracés + légendes
\newcommand{\valP} {1  };
\newcommand{\valPY}{0.5};
\newcommand{\valPZ}{  1};
\newcommand{\valPE}{  1};

\addplot[optimal,domain=1/(1+\valP/\valPY):1] {
	1/( 1/\valPE + max(1 , 1/x * (1+\valP/\valPZ)/(1+\valP/\valPY)) )
};
\addlegendentry{Optimal};

\addplot[digital] coordinates{
	(0.33333,0.26795)(0.34722,0.27911)(0.36111,0.29028)(0.375,0.30144)(0.38889,0.31261)(0.40278,0.32377)(0.41667,0.33494)(0.43056,0.3461)(0.44444,0.35727)(0.45833,0.36842)(0.47222,0.37931)(0.48611,0.38983)(0.5,0.4)(0.51389,0.40984)(0.52778,0.41935)(0.54167,0.42857)(0.55556,0.4375)(0.56944,0.44615)(0.58333,0.45455)(0.59722,0.46269)(0.61111,0.47059)(0.625,0.47826)(0.63889,0.48571)(0.65278,0.49296)(0.66667,0.5)(0.66667,0.5)(0.68056,0.5)(0.69444,0.5)(0.70833,0.5)(0.72222,0.5)(0.73611,0.5)(0.75,0.5)(0.76389,0.5)(0.77778,0.5)(0.79167,0.5)(0.80556,0.5)(0.81944,0.5)(0.83333,0.5)(0.84722,0.5)(0.86111,0.5)(0.875,0.5)(0.88889,0.5)(0.90278,0.5)(0.91667,0.5)(0.93056,0.5)(0.94444,0.5)(0.95833,0.5)(0.97222,0.5)(0.98611,0.5)(1,0.5)
};
\addlegendentry{Digital};

\addplot[analog,domain=1/(1+\valP/\valPY):1] {
	1/( 1 + 1/\valPE + (1/x - 1)*\valPY/\valPZ)
};
\addlegendentry{Analog};

\addplot[limites] coordinates{
	(1/3,0)(1/3,1)
};

\end{axis}
\end{tikzpicture}
\caption{Quantity $D_E$ as a function of the distortion level at Bob $D$ ($P=1$, $P_Y=0.5$, $P_Z=1$, $P_E=1$).}
\label{fig:DE=fD}
\end{figure}
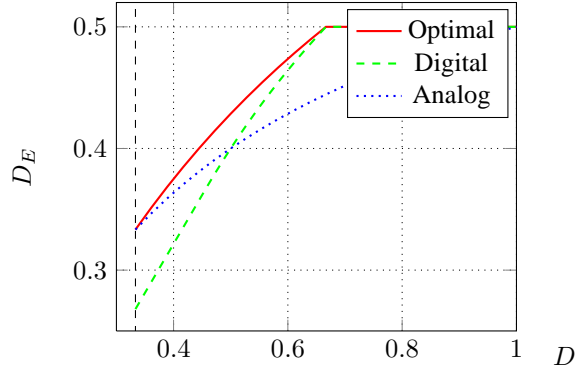

Fig.~\ref{fig:DE=fD} represents the largest achievable $D_E$ as a function of the distortion level at Bob $D$ for 
\begin{description}
\item[(i)] the optimal hybrid digital/analog scheme of Th.~\ref{th:gauss0Hybrid},
\item[(ii)] the digital scheme of Prop.~\ref{prop:gauss0Isit} (optimizing over $\mu$),
\item[(iii)] a pure analog scheme consisting of directly sending a scaled version of the source over the channel,
\end{description}
for parameter values $P=1$, $P_Y=0.5$, $P_Z=1$, $P_E=1$.

As a matter of fact, the proposed hybrid digital/analog scheme outperforms both pure analog and digital schemes.

\begin{remark}
While the analog scheme is optimal for $D = \frac1{1+\frac{P}{P_Y}}$ and, from Remark~\ref{rem:digitalOpt}, the digital one is optimal for $D \geq \frac{{1+\frac{P}{P_Z}}}{1+\frac{P}{P_Y}}$, a time-sharing combination of these falls short to achieve the entire region, as shown by Fig.~\ref{fig:DE=fD} and Theorem~\ref{th:gauss0Hybrid}.
\end{remark}

%::::::::::::::::::::::::::::::::::::::
%::::::::::::::::::::::::::::::::::::::

%==============================================================================
\section{Secure Transmission of a Binary Source over a Type-II Wiretap Channel}
\label{sec:binary}

%------------------------------------------------------------------------------
\subsection{System Model}

\begin{figure}[!b]
\centering
\begin{tikzpicture}
	\node	(ci)	at (-.4,0)	{};
	\node	(c0) 	at (0,0) 	{};
	\node	(c1)	at (2.5,0)	{};
	\node	(c2)	at (5,0)	{};
	\node	(c3)	at (8,0)	{};
	\node	(e) 	at (9,0) 	{};
	
	\draw[->] (ci) to (e);
	\draw (c0) to node[anchor=north] {\small $A\mkv B\mkv E$}		(c1);
	\draw (c1) to node[anchor=north] {\small $B\lessnoisy{A}E$}		(c2);
	\draw (c2) to node[anchor=north] {\small $I(A;B)\geq I(A;E)$}	(c3);
	
	\draw (c0)+(0,-2pt) -- +(0,2pt)	node[anchor=south]	{$0$};
	\draw (c1)+(0,-2pt) -- +(0,2pt)	node[anchor=south]	{$2\epsilon$};
	\draw (c2)+(0,-2pt) -- +(0,2pt) node[anchor=south] 	{$4\epsilon(1-\epsilon)$};
	\draw (c3)+(0,-2pt) -- +(0,2pt) node[anchor=south] 	{$h_2(\epsilon)$};
	\draw (e) node[anchor=south] 	{$\beta$};
\end{tikzpicture}
\caption{Relative properties of the side informations.}
\label{fig:cas}
\end{figure}
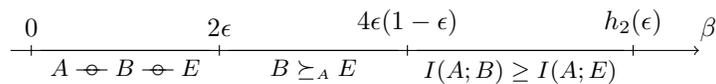

Consider now the binary example first proposed in~\cite{villard2011secureb}. In this setup, the source is binary uniformly distributed ($A\sim\cB\left(\tfrac12\right)$) and the side information at Bob, resp. Eve, is the output of a binary erasure channel (BEC) with erasure probability $\beta\in[0,1]$, resp. a binary symmetric channel (BSC) with crossover probability $\epsilon\in[0,\tfrac12]$, with input $A$.
According to the values of $(\beta,\epsilon)$ these side informations satisfy the properties summarized in Fig.~\ref{fig:cas}. 
The communication channel is similar to the one of~\cite{wyner1975wire}: It consists of a noiseless channel from Alice to Bob, and a BSC with crossover probability $\zeta\in[0,\tfrac12]$, from Alice to Eve. 
Let the distortion level at Bob be zero \emph{i.e.}, he performs \emph{lossless} reconstruction.

%::::::::::::::::::::::::::::::::::::::
%::::::::::::::::::::::::::::::::::::::

%------------------------------------------------------------------------------
\subsection{Performance of Coding Schemes}

Under the above assumptions, from~\cite[Proposition~3]{villard2011secureb}, the inner bound of Theorem~\ref{th:digital} is maximized by choosing $V=A$ and a uniformly distributed binary auxiliary RV $U$ (resp. $Q$), produced as the output of a BSC with crossover probability $u\in[0,\tfrac12]$ (resp. $q\in[0,\tfrac12]$), and input $A$ (resp. $X$). Consider also the hybrid scheme of Theorem~\ref{th:hybrid} choosing variables $U$, $V$ and $X$ as follows: $U = V \oplus W$, $V \stackrel{\indpt A}{\sim}\cB(\tfrac12)$ and $X = V \oplus A$ where $W$ is independent of $A$ and $V$, and $W\sim\cB(u)$ for some crossover probability $u\in[0,\tfrac12]$.

% - - - - - - - - - - - - - - - - - - - - - - - - - - - - - - - - - - - - - - -
\subsubsection*{Numerical Results}

\begin{figure}
\centering
\begin{tikzpicture}

% styles des différentes courbes
\tikzstyle{outer}	=[			color=black]
\tikzstyle{hybrid}	=[			color=red,	thick]
\tikzstyle{analog}	=[dotted,	color=blue,	thick]
\tikzstyle{digital}	=[dashed,	color=green,thick]
\tikzstyle{limites}	=[dashed,	color=black]

% limites et labels des axes
\newcommand{\xmin}{0}
\newcommand{\xmax}{1}
\newcommand{\ymin}{.12}
\newcommand{\ymax}{.495}

\newcommand{\xlabel}{$\beta$}
\newcommand{\ylabel}{$\Delta$}

\pgfplotsset{every axis x label/.append style={at={(1.12,.1)}}}

% - - - - - - - - - - - - - - - - - - - - - - - - - - - - - - - - - -
% axes, avec réglages de la grille, des légendes, des labels et des échelles pour abscisses et ordonnées
\begin{axis}[xlabel={\xlabel}, ylabel={\ylabel}, 
			xmin=\xmin, xmax=\xmax, ymin=\ymin, ymax=\ymax, 
			grid=both,
			width=6.85cm]

\pgfplotsset{every axis grid/.style={dotted}} 
\pgfplotsset{every axis legend/.append style={
	cells={anchor=west},
	at={(.03,.03)},
	anchor=south west
}}

% - - - - - - - - - - - - - - - - - - - - - - - - - - - - - - - - - -
% tracés + légendes
\newcommand{\valP} {1  };
\newcommand{\valPY}{0.5};
\newcommand{\valPZ}{  1};
\newcommand{\valPE}{  1};

\addplot[outer] coordinates{
	(0,0.469)(0.1,0.469)(0.2,0.469)(0.3,0.469)(0.36,0.469)(0.36,0.469)(0.41,0.469)(0.46,0.469)(0.469,0.469)(0.469,0.469)(0.479,0.46146)(0.489,0.45418)(0.499,0.44715)(0.509,0.44036)(0.519,0.4338)(0.529,0.42744)(0.539,0.42129)(0.549,0.41533)(0.559,0.40955)(0.569,0.40394)(0.579,0.3985)(0.589,0.39322)(0.599,0.38808)(0.609,0.3831)(0.619,0.37825)(0.629,0.37353)(0.639,0.36893)(0.649,0.36446)(0.659,0.36011)(0.669,0.35586)(0.679,0.35173)(0.689,0.34769)(0.699,0.34376)(0.709,0.33992)(0.719,0.33617)(0.729,0.33251)(0.739,0.32893)(0.749,0.32544)(0.759,0.32202)(0.769,0.31869)(0.779,0.31542)(0.789,0.31223)(0.799,0.3091)(0.809,0.30604)(0.819,0.30305)(0.829,0.30012)(0.839,0.29724)(0.849,0.29443)(0.859,0.29167)(0.869,0.28897)(0.879,0.28632)(0.889,0.28372)(0.899,0.28117)(0.909,0.27867)(0.919,0.27622)(0.929,0.27381)(0.939,0.27144)(0.949,0.26912)(0.959,0.26684)(0.969,0.26461)(0.979,0.26241)(0.989,0.26025)(0.999,0.25813)
};
\addlegendentry{Outer bound};

\addplot[hybrid] coordinates{
	(0,0.469)(0.1,0.469)(0.2,0.469)(0.3,0.469)(0.36,0.469)(0.36,0.469)(0.41,0.469)(0.46,0.469)(0.469,0.469)(0.469,0.469)(0.479,0.459)(0.489,0.449)(0.499,0.439)(0.509,0.429)(0.519,0.419)(0.529,0.409)(0.539,0.399)(0.549,0.389)(0.559,0.379)(0.569,0.369)(0.579,0.359)(0.589,0.349)(0.599,0.33932)(0.609,0.33046)(0.619,0.32238)(0.629,0.315)(0.639,0.30828)(0.649,0.30217)(0.659,0.29663)(0.669,0.29161)(0.679,0.28708)(0.689,0.28301)(0.699,0.27936)(0.709,0.2761)(0.719,0.27321)(0.729,0.27067)(0.739,0.26844)(0.749,0.2665)(0.759,0.26482)(0.769,0.2634)(0.779,0.26219)(0.789,0.26119)(0.799,0.26037)(0.809,0.2597)(0.819,0.25918)(0.829,0.25878)(0.839,0.25849)(0.849,0.25827)(0.859,0.25812)(0.869,0.25803)(0.879,0.25797)(0.889,0.25794)(0.899,0.25792)(0.909,0.25792)(0.919,0.25791)(0.929,0.25791)(0.939,0.25791)(0.949,0.25791)(0.959,0.25791)(0.969,0.25791)(0.979,0.25791)(0.989,0.25791)(0.999,0.25791)	
};
\addlegendentry{Hybrid};

\addplot[digital] coordinates{
	(0,0.469)(0.1,0.469)(0.2,0.469)(0.3,0.469)(0.36,0.469)(0.36,0.469)(0.41,0.469)(0.46,0.469)(0.469,0.469)(0.469,0.469)(0.479,0.459)(0.489,0.449)(0.499,0.439)(0.509,0.429)(0.519,0.419)(0.529,0.409)(0.539,0.399)(0.549,0.389)(0.559,0.379)(0.569,0.36906)(0.579,0.35938)(0.589,0.34995)(0.599,0.34077)(0.609,0.3318)(0.619,0.32304)(0.629,0.31447)(0.639,0.30607)(0.649,0.29784)(0.659,0.28976)(0.669,0.28182)(0.679,0.27401)(0.689,0.26633)(0.699,0.25876)(0.709,0.25129)(0.719,0.24393)(0.729,0.23665)(0.739,0.22946)(0.749,0.22235)(0.759,0.21531)(0.769,0.20833)(0.779,0.20142)(0.789,0.19456)(0.799,0.18775)(0.809,0.18099)(0.819,0.17428)(0.829,0.1676)(0.839,0.16095)(0.849,0.15434)(0.859,0.14775)(0.869,0.14119)(0.879,0.13464)(0.889,0.12812)(0.899,0.12161)(0.909,0.1151)(0.919,0.10861)(0.929,0.10213)(0.939,0.095642)(0.949,0.089159)(0.959,0.082675)(0.969,0.076187)(0.979,0.069693)(0.989,0.063192)(0.999,0.056681)
};
\addlegendentry{Digital};

\addplot[analog] coordinates{
	(0,0.25791)(0.1,0.25791)(0.2,0.25791)(0.3,0.25791)(0.36,0.25791)(0.36,0.25791)(0.41,0.25791)(0.46,0.25791)(0.469,0.25791)(0.469,0.25791)(0.479,0.25791)(0.489,0.25791)(0.499,0.25791)(0.509,0.25791)(0.519,0.25791)(0.529,0.25791)(0.539,0.25791)(0.549,0.25791)(0.559,0.25791)(0.569,0.25791)(0.579,0.25791)(0.589,0.25791)(0.599,0.25791)(0.609,0.25791)(0.619,0.25791)(0.629,0.25791)(0.639,0.25791)(0.649,0.25791)(0.659,0.25791)(0.669,0.25791)(0.679,0.25791)(0.689,0.25791)(0.699,0.25791)(0.709,0.25791)(0.719,0.25791)(0.729,0.25791)(0.739,0.25791)(0.749,0.25791)(0.759,0.25791)(0.769,0.25791)(0.779,0.25791)(0.789,0.25791)(0.799,0.25791)(0.809,0.25791)(0.819,0.25791)(0.829,0.25791)(0.839,0.25791)(0.849,0.25791)(0.859,0.25791)(0.869,0.25791)(0.879,0.25791)(0.889,0.25791)(0.899,0.25791)(0.909,0.25791)(0.919,0.25791)(0.929,0.25791)(0.939,0.25791)(0.949,0.25791)(0.959,0.25791)(0.969,0.25791)(0.979,0.25791)(0.989,0.25791)(0.999,0.25791)
};
\addlegendentry{Analog};

\addplot[limites] coordinates{
	(0.36,0)(0.36,1)
};
\addplot[limites] coordinates{
	(0.469,0)(0.469,1)
};

\end{axis}
\end{tikzpicture}
\caption{Equivocation rate at Eve $\Delta$ as a function of the erasure probability $\beta$ ($\epsilon=0.1$, $\zeta=0.1$).}
\label{fig:Delta=fbeta}
\end{figure}
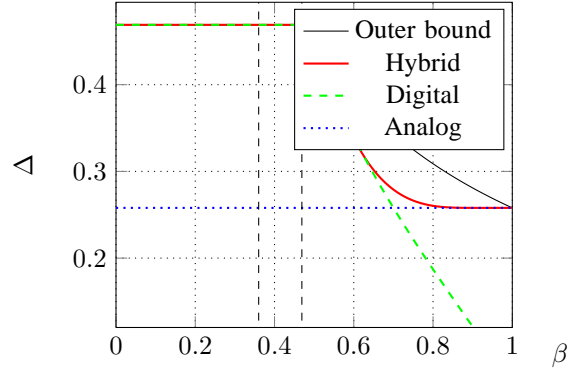

Fig.~\ref{fig:Delta=fbeta} represents the equivocation rate $\Delta$ as a function of the erasure probability $\beta$ for 
\begin{description}
\item[(i)] the outer bound \emph{i.e.}, Theorem~\ref{th:outer},
\item[(ii)] the hybrid digital/analog scheme of Theorem~\ref{th:hybrid} with $U$, $V$, $X$ as defined above (and optimizing over $u$),
\item[(iii)] the digital scheme of~\cite[Proposition~3]{villard2011secureb}, optimizing over $u$ and $q$,
\item[(iv)] a pure analog scheme consisting of directly sending the source over the channel,
\end{description}
for parameter values $\epsilon=0.1$, $\zeta=0.1$.

As expected, if $\beta\leq 4\epsilon(1-\epsilon)$, $B$ is less noisy than $E$, and the digital scheme is optimal, as well as the proposed hybrid one.
Here, this result also seems to hold when $B$ is only more capable than $E$ \emph{i.e.}, for $\beta\leq h_2(\epsilon)$.

For $\beta=1$, as noted in~\cite{villard2011secureb}, the naive pure analog scheme outperforms the digital one.
According to the comments of Section~\ref{sec:specAnalog}, the proposed hybrid digital/analog scheme always performs as good as the analog one. In Fig.~\ref{fig:Delta=fbeta}, the proposed hybrid digital/analog scheme also seems to perform as good as the digital one. However, according to the comments of Section~\ref{sec:specDigital}, and depending on the parameters $\epsilon$, $\zeta$, this may not be the case for all values in $\left[ h_2(\epsilon), 1 \right)$.
 
%::::::::::::::::::::::::::::::::::::::
%::::::::::::::::::::::::::::::::::::::

%%%%%%%%%%%%%%%%%%%%%%%%%%%%%%%%%%%%%%%%%%%%%%%%%%%%%%%%%%%%%%%%%%%%%%%%%%%%%%%
%%%%%%%%%%%%%%%%%%%%%%%%%%%%%%%%%%%%%%%%%%%%%%%%%%%%%%%%%%%%%%%%%%%%%%%%%%%%%%%
\bibliographystyle{IEEEtran}
\bibliography{itw2011}

\end{document}